\documentstyle[preprint,revtex]{aps}
\begin{document}
\draft
\begin{title}
The nature of the highest energy cosmic rays
\end{title}
\author{Todor Stanev$^1$ \& H.P. Vankov}
\begin{instit}
$^1$ Bartol Research Institute, University of Delaware, Newark, DE 19716\\
$^2$ Institute for Nuclear Research and Nuclear Energy, Sofia \#\#, Bulgaria\\
\end{instit}
\begin{abstract}
 Ultra high energy gamma rays produce electron--positron pairs  in
 interactions on the geomagnetic field. The pair electrons suffer
 magnetic bremsstrahlung  and the energy of the primary gamma ray is
 shared by a bunch of lower energy secondaries. These processes
 reflect the structure of the geomagnetic field and cause experimentally
 observable effects. The study of these effects with future giant
 air shower arrays can identify the nature of the highest energy
 cosmic rays as either $\gamma$--rays or nuclei.
\end{abstract}


\section{Introduction}

  Ever since the reports of the detection of two cosmic ray showers
 of energy well above $10^{20}$ eV~\cite{FEPRL,AKH} the origin
 and the nature of such events have been subjects of strong interest
 and intense discussion. It is not only very difficult~\cite{HillAnnRev}
 to extend our understanding of particle acceleration to such 
 extraordinarily high energies but the propagation of these particles
 in the microwave background and possibly other universal radiation
 fields restricts the distance to their potential sources to several
 tens of Mpc.

  Conservatively minded astrophysicists are looking for astrophysical
 sources which may contain the environment necessary for stochastic
 particle acceleration to energies in excess of $10^{20}$ eV. 
 Powerful (FRII) radio galaxies~\cite{RB12} have been suggested as
 possible sources. If this suggestion were true, the highest energy cosmic
 rays (HECR) would be most likely protons, reflecting the composition
 of the matter that is available for injection in the termination
 shocks of FRII jets. Others~\cite{Arnold} search for powerful
 astrophysical sources in the cosmologically nearby Universe. HECR
 then could also be heavier nuclei, for which the acceleration
 is less demanding. The propagation of heavy nuclei on short distances
 (O(10) Mpc) without huge energy loss is possible.

  Some cosmologists relate the origin of HECR to topological
 defects~\cite{BhHSSL}. Topological defects (TD) scenarios avoid
 the problems of  particle acceleration since they are based on
 `top--down' evolution. Very massive ($10^{22} - 10^{25}$ eV)
 X--particles are emitted by the topological defects that later decay
 into baryons and mesons of lower energy. Most of the energy is
 eventually carried by $\gamma$--rays and neutrinos, that are products
 of meson decay. Detected HECR would then most likely be $\gamma$--rays. 

  Most radically, the origin of HECR has been related to those
 of gamma ray bursts~\cite{GRB1,GRB2,GRB3}, replacing two extremely luminous
 mysteries with a single one. In such scenarios  HECR are most likely
 to be again protons. We may not be able to observe the sources of
 HECR since every source might only emit a single  observed ultrahigh
 energy particle. 
 
  The nature, the type of the particle that interacted in the
 atmosphere to generate these giant air showers, could be the key
 to the understanding the origin of the highest energy cosmic
 rays.  The current experimental evidence on the nature of HECR is
 not conclusive.  The Fly's Eye experiment, for example, has reported
 correlated changes in the spectra and the composition of the ultra high
 energy cosmic rays~\cite{FYSC}. The analysis of the Fly's Eye
 experimental statistics suggests that a change of the chemical
 composition of the cosmic rays  from heavy nuclei to protons at
 $\sim 3 \times 10^{18}$ eV is accompanied by a change of the
 spectral index of the cosmic ray energy spectrum. One may then
 conclude that the HECR are protons. The other currently running
 air shower experiment, AGASA,  does not observe~\cite{AGASAS}
 such a correlation. A re--analysis of the archival data from the 
 SUGAR experiment~\cite{WWSUGAR} makes the opposite conclusion -- 
 a large fraction of the highest energy showers seem to be generated
 by heavy nuclei..

  A correlation between the arrival directions of HECR with energy 
 $> 4 \times 10^{19}$ eV with the supergalactic plane,
 that contains most of the galaxies of redshift $<$ 0.03, has been
 reported~\cite{PRL}. The AGASA experiment has also observed similar
 correlation in their data set~\cite{ASGP}, although not fully
 consistent with the conclusions of Ref.~\cite{PRL}. On the other hand
 the Fly's Eye experiment does not see such a correlation (P. Sommers
 for the Fly's  Eye group, {\it private communication}). It also has 
 not been observed in the SUGAR data~\cite{Adel}.  Even if confirmed
 in the future, a  correlation with the structure of the local universe
 would not answer the question of the nature of HECR. If topological
 defects are seeds for galaxy formation most powerful galaxies and TD
 would have similar distribution and  TD and astrophysical scenarios
 of the origin of HECR are indistinguishable.   
  
 The profile of the 3$\times 10^{20}$ eV  shower detected by the
 Fly's Eye develops higher in the atmosphere than expected for
 either proton or $\gamma$--ray showers of that energy~\cite{HSV}.
 The highest energy shower seen by the AGASA experiment ($2
 \times 10^{20}$ eV) exhibits, apart from its energy, features that
 are typical for most of the high energy showers. The currently
 existing air shower arrays cannot drastically increase the
 experimental statistics and the hope for answering
 the important questions for the nature and origin of HECR is in
 the construction of much bigger shower arrays, such as the Auger
 project~\cite{Auger}. 

  Even with Auger, however, the nature of HECR will be difficult
 to study. Shower parameters are subject of strong intrinsic
 fluctuations and the cross sections that govern inelastic
 interactions at $\sqrt{s}$ = 100 TeV are not well enough known.
 At lower energy ($10^{14} - 10^{16}$ eV) showers generated by
 heavy nuclei, protons and $\gamma$--rays could be at least
 statistically distinguished by their muon content. $\gamma$--ray
 showers have on the average $\sim 3$\% of the muon content of
 proton showers of the same energy~\cite{GSH?}. At ultrahigh
 energies such approach may not be possible --  calculations of
 the muon content of the $\gamma$--ray induced showers predict that
 the fraction of GeV muons  could be even higher than in proton
 generated showers~\cite{TP,Ahhh}.
 
  We suggest a different approach to the study of the nature of
 the cosmic rays with energy above $10^{19}$ eV -- to prove (or disprove)
 that HECR are $\gamma$--rays by observing their interactions with
 the geomagnetic field. While protons and heavier nuclei are not
 affected by the geomagnetic field, ultra high energy $\gamma$--rays
 interact on it to produce $e^+e^-$ pairs. The electrons themselves
 quickly lose their energy through magnetic bremsstrahlung (synchrotron
 radiation) before they enter the atmosphere of the earth. Air  showers
 are thus replaced by  `magnetic + atmospheric' showers that start
 far away from the  surface of the earth and are absorbed faster
 compared to usual air showers. With high experimental statistics one
 can observe the interactions of ultra high energy $\gamma$--rays with
 the geomagnetic field by a study of the shower arrival direction in
 geographical coordinates. If the detected showers do not show
 signs of interactions with the geomagnetic field, the suggestions
 for $\gamma$--ray nature of HECR could be proven wrong.

  This article is organized in the following way. Section 2. gives
 a brief discussion of the photon and electron interactions on 
 magnetic fields and of the structure of the geomagnetic field.
 Section 3. describes a calculation of the `geomagnetic + atmospheric'
 cascades and gives some general results of that calculation.
 Section 4. calculates shower parameters that could be used to
 confirm the $\gamma$--ray origin of HECR and Section 5. contains
 the conclusions from this research.
 
\section{Photon and electron interactions in the geomagnetic field}

  Interactions of photons, and especially of electrons, on
 magnetic fields have been exhaustively studied because of
 all the problems they create in particle accelerators. 
 The theoretical and some experimental knowledge is reviewed
 by T. Erber in Ref.~\cite{Erber}. 

  Magnetic pair production is guided by the parameter
 $\Upsilon_\gamma \equiv [1/2] [h\nu/mc^2] [B_\perp/B_{cr}]$, where
 $B_{cr} \equiv m^2c^3/e \hbar = 4.414 \times 10^{13}$ Gauss and
 $B_\perp$ is the component of the magnetic field that is 
 normal to the $\gamma$--ray trajectory. The $\gamma$--ray
 attenuation coefficient, i.e. the fraction of photons that
 undergo pair production in magnetic field of strength $B_\perp$
 per unit distance is given by
\begin{equation}
 \alpha_\gamma (\Upsilon_\gamma)\; = \; 0.16 
{{\alpha } m c\over {\hbar}} {{mc^2} \over {h \nu}}
K^2_{1/3} (2\Upsilon_\gamma/3) {\rm cm}^{-1}
\label{attp}
\end{equation}
 The maximum attenuation is reached at $\gamma$--ray energy
 of $12 m c^2 (B_{cr}/B_\perp)$ while the cross section of the process
 is linearly proportional to the magnetic field strength $B_\perp$. 
  
  Similarly the magnetic bremsstrahlung (synchrotron radiation) 
 is guided by $\Upsilon_e \equiv [E/m c^2] [ B_\perp/B_{cr}]$.
 The radiation emitted by an electron of energy $E_e$ in magnetic field
 $B_\perp$ per unit distance is distributed as
\begin{equation}
 I(E_e, h\nu, B_\perp)\;=\; {{sqrt{3}\alpha}\over{2\pi}} {{ m^2 c^3} \over
 {\hbar}}  {{\Upsilon_e} \over {E}} 
 \left( 1 - {{h\nu}\over{E_e}}\right) K ( 2 J )\;,
\label{synch}
\end{equation}
 where $J \equiv [h\nu/E_e][1 + h\nu/E_e]/3\Upsilon_e$. 

  To demonstrate the strength of the  $\gamma$--ray interactions
 in the geomagnetic field we show in Fig.~\ref{fig1} the distributions
 of the distances from the surface of the earth at which $\gamma$--rays
 of different energy pair produce. The $\gamma$--ray trajectory is
 taken to be normal to the field lines of a magnetic dipole centered
 at the center of the Earth with magnetic moment of $8.1 \times 10^{19}$
 Gauss/m. One could see that the $\gamma$--rays of the energies of
 interest interact in a relatively narrow range of distances not
 further than 3$R_\oplus$.  The narrow peak plotted at altitude of 20 km
 represents $\gamma$--rays that survive, i.e. interact in the atmosphere
 before they interact in the geomagnetic field. 12\% of the
 $\gamma$--rays with energy $10^{20}$ eV (and none at higher energy)
 survive.

  The spectra of the $\gamma$--rays emitted in magnetic bremsstrahlung
 depends quite strongly on the magnetic field strength. For strong fields
 the energy distribution of the secondary photons is quite flat.
 Fig.~\ref{fig2} shows the energy loss of  $10^{20}$ electrons in
 magnetic fields of strength  $\log_{10}{B_\perp}$ = --0.5, --1, --1.5, etc.
 Gauss as a function of the secondary photon energy. In the dipole
 field model described above a field of 0.1 Gauss corresponds to a
 distance of 0.468 $R_\oplus$ above the surface of the earth, and
 0.032 Gauss -- to 1.15 $R_\oplus$. These distances cover much of
 the primary $\gamma$--rays interaction range shown in Fig.~\ref{fig1}.
 Since lower energy $\gamma$--rays  pair produce close to the earth,
 the magnetic bremsstrahlung of their secondary electrons is harder.
 The energy spectra of the $\gamma$--rays in the bunch that enters the
 atmosphere after pair production and magnetic bremsstrahlung tend
 to be almost independent of the primary $\gamma$--ray energy $E^0_\gamma$.

\subsection{Structure of the geomagnetic field}

 Gamma rays arriving at any experimental location under different
 zenith ($\vartheta$) and azimuthal ($\phi$) angle will see
 a different geomagnetic field. They will thus cascade differently
 before reaching the atmosphere. At small $\vartheta$, close to the
 vertical direction, the variation with $\phi$ is insignificant.
 At relatively large $\vartheta$, more than 30$^\circ$, the field
 strength for most locations changes by factors of 3 or more for
 different values of $\phi$.

  A more quantitative calculation of the strength of the field
 encountered by the incoming $\gamma$--ray is trivial for any model
 of the geomagnetic field but has to be performed for each location,
 $\vartheta$ and $\phi$ separately. We have attempted to obtain a
 slightly more general result for several experimental locations.
 Fig.~\ref{fig3} shows the transverse component ($B_\perp$) of the
 geomagnetic field as a function of the azimuthal angle at which
 it arrives to the detector. Since the $\phi$ variations for
 different zenith angles $\vartheta$  have the same aspect, we have
 integrated over $\vartheta$ from 0 to 60$^\circ$, weighting the
 field values with the solid angle. The 1991 International Geomagnetic 
 Reference Field  model (IGRF) is used for this calculation. 

  Four of the locations for which $B_\perp$ is shown
 are in the Northern hemisphere and only one (Sydney, 
 shown with dash--dash) is South of the equator.
 Since the smallest $B_\perp$ is seen in the direction of the
 magnetic pole that is closer to each location, northern and
 southern locations have opposite field strength dependence
 on $\phi$. At the moderate latitudes of these detector
 locations, the detailed differences between the Northern
 hemisphere detectors are minor. (A detector located at
 the geomagnetic equator would have a symmetric response
 to geomagnetic North and South directions). The difference
 between the maximum and minimum field strengths is 
 almost a factor of 5.

  For each one of these detectors, as well as for any other detector
 location, one could determine a region in azimuth, where the
 field strength is the lowest and the incoming $\gamma$--rays would
 be affected minimally by the geomagnetic field and a region where
 the effect of the geomagnetic field is at maximum. For the location
 of Sydney, e.g. $\gamma$--rays arriving with $130^\circ < \phi <
 215^\circ$ would see $B_\perp <$0.02 Gauss at a distance of 1 $R_\oplus$ 
 and $\gamma$--rays with $ 255^\circ < \phi < 90^\circ$ would
 see more than 0.04 Gauss at the same distance. The idea is that
 $\gamma$--ray fluxes arriving from these two regions may have
 observable characteristics that are different enough to be 
 distinguished experimentally. We continue to study
 the cascading of ultra high energy gamma rays in geomagnetic fields
 with different strength, corresponding to these two regions.
  
\section{Cascading in the geomagnetic field}

  We simulate the electromagnetic cascading in the geomagnetic field 
 by injecting  $\gamma$--rays of energy $E^0_\gamma$ at a distance of 5
 $R_\oplus$ from the surface of the earth on a trajectory with angle
 $\vartheta$ relative to the vertical direction at the intersection with
 the surface. The $\gamma$--ray is propagated with a stepsize $\Delta x$
 (from 1 to 10 km) until the $\gamma$--ray pair produces or reaches
 the atmosphere. The atmosphere is defined to be at altitude of
 20 km above the earth's surface. Gamma rays that reach the
 atmosphere `survive' and interact in the atmosphere to produce
 air showers with their original injection energy.

  If the $\gamma$--ray produces an electron--positron pair, the
 pair electrons are followed in similar way, by calculating their
 radiation spectrum on every step of propagation. The synchrotron
 $\gamma$--rays are tabulated in energy, starting at $10^{14}$ eV.
 The assumption here is that secondary $\gamma$--rays of energy less
 than $10^{14}$ eV do not contribute significantly to the
 cascades that are observed deep in the atmosphere. This lower energy
 end of the magnetic bremsstrahlung spectrum, as well as the electrons
 of energy below $10^{14}$ eV that enter the atmosphere, contain always
 less than 2\% of the primary $\gamma$--ray.

  Each particle produced in the geomagnetic field, as well as the
 `surviving' primary $\gamma$--rays then generate atmospheric 
 cascades. The profiles of these cascades are added up to calculate
 the composite shower profile, generated in the atmosphere by the
 injected primary $\gamma$--ray or the products of its interaction 
 in the geomagnetic field.
 
  The actual calculation is performed using the dipole magnetic field
 model with magnetic moment of $8.1 \times 10^{19}$ Gauss/m with two
 scale factors of 0.25 (low field) and 1.25 (high field), which generate
 magnetic field strengths approximately equal to the maximum and
 minimum values shown in Fig.~\ref{fig3}.
 
  To study the `survival' probability in these two field models we
 made calculations for two extreme zenith angles: $\vartheta$ = 0$^\circ$
 and 60$^\circ$, which is the maximum zenith angle at which air showers
 can be reliably detected and analyzed. Fig.~\ref{fig4} shows the
 survival probabilities at high (lefthand strip) and low
 magnetic field strengths. The lefthand boundary of each strip
 corresponds to propagation at $\vartheta$ = 60$^\circ$ and the righthand
 boundary is for $\vartheta$ = 0$^\circ$. $\gamma$--rays approaching
 the earth at higher zenith angles spend significantly more time 
 in higher geomagnetic field strengths and have a higher interaction
 probability. The lefthand edge of the high field strip and the
 righthand edge of the low field strip practically bracket the
 survival probability space for $\gamma$--rays approaching any location
 at the earth surface with zenith angles smaller than 60$^\circ$. 
 $\gamma$--rays arriving at higher angles may be absorbed faster.

 Several calculations of the $\gamma$--ray cascading in the geomagnetic
 field have been previously performed~\cite{McBLa,Ahhh,VS}. Our results
 are in a good agreement with the main results of all of them.
 Our calculation is  generally a refinement of previous ones, which
 nevertheless reveals some practically important features in the
 cascading process. Previous calculations conclude that there will
 be a `cutoff' in the energy spectrum of the $\gamma$--rays that
 reach the atmosphere, because of the very soft spectrum of the
 secondary photons, generated by magnetic bremsstrahlung. This
 conclusion is partially due to the relatively rough treatment and
 low statistics in the previous work.
 Fig.~\ref{fig5} shows the number of secondary $\gamma$--rays and the
 energy  that they carry. $E^0_\gamma$ = $3 \times 10^{20}$ eV in this
 example. Although the number of secondary $\gamma$--rays of energy
 above $10^{19}$ eV is on the average only 6.3, they carry 48\% of
 the primary energy. This is also important for the development of
 the subsequent air showers, because at energies above $10^{19}$ eV
 the LPM effect~\cite{LPM}, which  suppresses the electromagnetic
 cross sections at high energy and slows  the development of the
 air showers, becomes important in air.

\section{Atmospheric showers}

  Gamma rays of energy above $10^{19}$ eV, if they do exist,
 would only be detectable by giant air shower arrays located
 on the surface of the earth. Air shower arrays consist of a large
 number of counters that trigger in coincidence when the shower
 from arrives. The shower direction is determined by the arrival
 time of the shower front at the different counters. A fit of the
 density in the separate counters reconstructs the total number
 of shower particles, the showers size $N_e$, which is then used to
 determine the primary energy.

   The output of our  Monte Carlo simulation includes the shower
 sizes calculated for several atmospheric depths from the cascading
 of all secondary (and primary, if the injected gamma rays did not
 pair produce) $\gamma$--rays in the atmosphere. The profiles 
 from individual secondary $\gamma$--rays of energy above $10^{18}$ eV
 are calculated with an account for the LPM effect, although the
 effect is not significant below $10^{19}$ eV. The depths are
 arbitrarily chosen to include a realistic range for typical large
 air shower experiment and correspond to an array at vertical
 depth of 860 g/cm$^2$ and zenith angles with $\cos{\vartheta}$ = 0.9,
 0.8, 0.7, 0.6 and 0.5. Because the development of purely electromagnetic
 showers in the atmosphere (apart from their muon content) does not
 depend significantly on the atmospheric density profile, the examples
 given below could be scaled also to different altitudes and zenith 
 angles.  
 
  Fig.~\ref{fig6} shows a general and important shower parameter --
 the average size ($<N_e>$) generated by $\gamma$--rays of
 different energy. The solid lines are for low field (scale factor of
 0.25) and the dashed lines are for high field (scale factor
 of 1.25). From top to bottom the lines show $N_e(E^0_\gamma)$ at
 the 5 different depths of 956, 1075, 1229, 1433 and 1720 g/cm$^2$.
 Except for the deepest observation level,  $N_e$ is
 multiplied by the factor shown by each curve to make the figure
 readable. In the absence of interactions in the geomagnetic field,
 and for lower $E^0_\gamma$, the shower size  have a power low
 dependence on $E_\gamma$, $N_e\;=\;E_\gamma^\alpha$ with $\alpha >$1.
 The power law index $\alpha$ depends on the column density between
 the depth of the shower maximum $X_{max}$ and the detector. The
 size at maximum $N_{max}$ is exactly proportional to $E^0_\gamma$
 and  $\alpha$ is bigger than unity because the depth of shower
 maximum grows with energy as
 $X_{max} = \log({E^0_\gamma/{\rm 81 MeV}})$ radiation lengths
 (1 r.l. = 37.1 g/cm$^2$ in air).

  The dependence shown in Fig.~\ref{fig6} is more complicated because
 in this energy range showers are already at or before their maximum
 development at some of the shallower observation levels. The role
 of the magnetic field strength on the $N_e$ dependence on $E^0_\gamma$
 is easier to understand for the deepest levels of observation. 
 Compare, for example, the two curves for depth of 1720 g/cm$^2$ with
 the $\gamma$--ray survival probability of Fig.~\ref{fig4}.
 At low energy, where there are no interactions on the geomagnetic
 field, the two curves are the same.  The solid
 curve (low field) starts bending at $\gamma$--ray energy 
 $2 \times 10^{20}$ eV where the primary $\gamma$--rays start 
 interacting in the geomagnetic field. Because of these interactions
 the primary $\gamma$--ray is replaced by a bunch of $\gamma$--rays
 of lower energy. The composite shower reaches maximum at shallower
 atmospheric depths and is significantly absorbed at the deep
 observation level. The same happens at energy lower by about one
 order of magnitude in the high field case.  Although it is outside
 of the energy range of Fig.~\ref{fig6}, at some higher energy, where 
 all $\gamma$--rays interact on the geomagnetic field, the two curves
 will join again. 

  To explain the behaviour at the shallow observation levels one has
 to take into account some of the details of the cascading in
 the geomagnetic field, namely the shape of the energy spectra of the
 secondary photons as a function of field strength, which is shown in
 Fig.~\ref{fig2}. Although the primary $\gamma$--rays interact in the
 same way, in the high field case the energy spectra of the
 secondary $\gamma$--rays are harder, hard enough to generate showers
 that are not absorbed at the level of 956 g/cm$^2$. One could hardly
 see a tiny deviation of the strong field (dashed) curve in the
 region of $E^0_\gamma = 3 \times 10^{19}$ eV. At higher energies 
 the secondaries are energetic enough to produce $N_e$ dependence
 very close to a power law. When the primary $\gamma$--rays start
 interacting in the low field, however, the picture is slightly
 different. The secondary $\gamma$--ray spectra are softer, the composite 
 showers reach maxima at shallower depths and are correspondingly 
 absorbed when they reach the observation level. The two curves will
 join asymptotically.

  All other levels show intermediate behaviour where the relation
 between the depth of observation and $X_{max}$ also contributes
 to the exact shape of the curve.

  Fig.~\ref{fig6} shows the strong differences in the observable
 parameter $N_e$ which is introduced by the strength of the
 geomagnetic field. It can not be used, however, for analysis
 of experimental data because $E^0_\gamma$ is not a directly
 measurable parameter. What experiments can do, and usually do,
 is to produce a spectrum of the measured shower sizes $N_e$.
 Such spectra for the three deeper observation levels are shown
 in Fig.~\ref{fig7}. The solid histogram corresponds to the low field
 and the dashed one -- to the high field case.

  The histograms are result of a simulation, where $E^0_\gamma$ 
 is sampled from a $(E^0_\gamma)^{-2}$ differential primary
 spectrum between $10^{19}$ and $10^{21}$ eV. At low $N_e$
 the spectra are always higher for the low field case, including
 the two observation levels that are not shown in Fig.~\ref{fig7}.
 At the high $N_e$ side and for shallow observation levels the
 high field case shows higher spectrum, as could be expected by
 the results shown in Fig.~\ref{fig6} and as seen for the shallowest
 level plotted in Fig.~\ref{fig7}. The biggest difference is at the
 deepest observation level, where the spectra are different as much
 as a factor of 10. The differences between the size spectra
 decreases for shallower observation levels, and is probably not
 detectable for the two shallowest levels, which are not shown. 

\section{Discussion and conclusions}

  The size spectra of Fig.~\ref{fig7} show that it is possible
 to detect the difference between a flux of $\gamma$--rays that reach
 the earth after cascading in geomagnetic field of different
 effective strength. In practical terms this means that any
 experiment that is able to collect large enough experimental
 statistics should see different $N_e$ spectra in different
 azimuthal directions if HECR are indeed $\gamma$--rays.
 We have not attempted to look for this effect in the existing
 experimental statistics, because it is not large enough to
 reveal such effects.

  The Auger project~\cite{Auger} is an entirely different story.
 It proposes the construction of two air shower arrays, at least
 3 000 km$^2$ each, in the Northern and in the Southern hemisphere.
 For comparison, the area of the largest current detector (AGASA)
 is 100 km$^2$. Each one of these detectors will have the collecting
 power of more than 5 000 showers above $10^{19}$ eV per year.
 If the HECR primaries are $\gamma$--rays the Norther hemisphere
 detector should see spectra similar to the low field case of
 Fig.~\ref{fig7} in northern direction (the exact direction and
 value of the minimum field depends on the location of the array)
 and the strong field case in showers coming from South. The southern
 detector will have the opposite effect. If this were the case, the
 difference of the size spectra  would prove that the HECR are gamma rays.

  The actual effects could even be stronger than shown in Fig.~\ref{fig7},
 because the simulation on which it is based propagated all
 primary $\gamma$--rays along the vertical ($\vartheta$ = 0$^\circ$)
 direction. In fact, as shown in Fig.~\ref{fig4}, the increase of the
 interaction probability in the geomagnetic field increases by a 
 non--negligible factor when the exact particle trajectory is
 accounted for. Our air shower simulation also does not account 
 for the magnetic bremsstrahlung of the shower electrons which
 at high $E_e$ and low atmospheric density $< 10^{-5}$ g/cm$^3$
 could be important and could accelerate the shower development.

  In principle the interplanetary magnetic field has to be added
 to the `target' magnetic field. Gamma ray arriving from a cone
 centered on the Sun would be absorbed far away from the earth and
 possibly not detectable. The sun could thus be visibly in ultra
 high energy $\gamma$-rays.  The exact dimensions of the region
 where $\gamma$--rays are absorbed in pair production on the solar
 magnetic field, carries valuable  information on the magnetic field
 in the vicinity of the sun. This is an interesting although purely
 academic problem, because the statistics of such events is always going
 to be negligible.

  Although we have not done it for this paper, there will be effects,
 similar to the $N_e$ ones, on the muon content of the $\gamma$--ray
 initiated air showers. It is well known that at these extremely
 high energies the number of soft muons (0.3 to 2 GeV) in $\gamma$
 initiated showers is comparable to this of hadronic showers~\cite{TP,Ahhh}.
 The number of soft muons has an $E^0_\gamma$ dependence very similar
 to $N_e$, because  the low energy muons decay readily when $X_{max}$
 is distant from the observation level. The decay length of 1 GeV
 muons is $\sim$6 km. A picture similar to the $N_e$ spectra in
 Fig.~\ref{fig7} will develop as a result of the cascading in 
 geomagnetic fields of different effective strength.
 The major difference between the behaviour of the electron size 
 and the muon size is that $N_e$ attenuates as a function
 of the column density, while  $N_\mu$ attenuates as a
 function of the distance, i.e. $N_\mu$ will depend strongly
 on the shower zenith angle $\vartheta$.     
  
  The study of the two major components of the giant air showers can
 reveal the nature of the highest energy cosmic rays. If the 
 specific dependence on the shower arrival direction is observed,
 then the highest energy cosmic rays are $\gamma$--rays. A non
 observation of this effect would leave us with the choice between
 protons and heavy nuclei.
 
 {\bf Acknowledgements.} TS is grateful to A.A. Watson for inspiring
 discussions on the subject of giant air shower. The work of TS is
 supported in part by the U.S. Department of Energy under contract
 DE-FG02-91ER40626. HPV is thankful to US NSF for partial support 
 of his visit to U.S. where this work was conceived, and to the Bartol
 Research Institute for its hospitality.

\figure{ Distribution of the interaction points of $\gamma$--rays of
 energy $10^{21}$ eV (solid line), $3.16 \times 10^{20}$ eV (dotted line)
 and $10^{20}$ eV (dashed line). The interaction points are the vertical
 distances from the surface of the earth. The shading in the left hand side
 of the figure represents the atmosphere.  
\label{fig1}
}
\figure{ Energy loss of $10^{20}$ eV electrons as a function of
 the strength of the magnetic field and the energy of the secondary
 photons. The field strength is indicated by the respective
 curve as $\log_{10} (B_\perp/{\rm Gauss})$.  
\label{fig2}
}
\figure{ The strength of the geomagnetic field component that
 is perpendicular to the $\gamma$--ray trajectory as a function
 of the azimuthal angle $\phi$ at which the particle arrives at
 the location is shown for a distance of 1 $R_\oplus$ from
 the detector. The field strength is integrated over zenith
 angles $\vartheta$ from 0 to 60$^\circ$ accounting for the solid 
 angle.  The calculation is performed for the locations
 of several air shower arrays: a.) Fly's Eye (40N, 112W) -- solid line;
 b.) Yakutsk (62N, 129E) -- dots; c.) Akeno (35N, 138E) -- dashes;
 d.) Haverah Park (54N,2W) -- dash--dot; e.) Sydney (30S, 150E) -- dash--dash.
 The 1991 IGRF model of the geomagnetic field is used in this calculation.
\label{fig3}
}
\figure{ Survival probability for $\gamma$--rays of energy between
 $10^{19}$ and $10^{21}$ eV in the dipole geomagnetic field model
 described in the text with scaling factors of 0.25 (righthand strip)
 and 1.25 (lefthand strip). The lefthand edge of each strip shows
 the survival probability for $\gamma$--rays approaching the surface
 of the earth with a zenith angle $\vartheta$ = 60$^\circ$ and the
 righthand edges are for $\vartheta$ = 0$^\circ$.
\label{fig4}
}
\figure{ Energy  spectrum of the secondary $\gamma$--rays that reach
 the atmosphere after the cascading of a primary $\gamma$--ray of
 energy $3\times10^{20}$ eV -- solid line, lefthand scale. The
 dotted histogram and the righthand scale show the amount of energy
 carried by the secondary $\gamma$--rays in each bin.
\label{fig5}
}
\figure{ Relation between the average shower size $N_e$ and the
 primary $\gamma$--ray energy $E^0_\gamma$ for the five observation
 levels defined in the text for cascading in high (solid) and low
 (dashed line) strengths of the geomagnetic field. $N_e$ values
 are multiplied by the factor indicated by the curves. 
\label{fig6}
}
\figure{ Integral shower size $N_e$ spectra generated by primary
 $\gamma$--rays sampled on a $(E^0_\gamma)^{-2}$ differential
 spectrum between $10^{19}$ and $10^{21}$ eV. The solid histograms
 show the low field case and the dashed histograms are for high field.
 The observation levels are 1720, 1433, and 1229 g/cm$^2$ from
 left to right.  
\label{fig7}
}

\begin{references}
%
\bibitem{FEPRL} D.J.~Bird {\em et al., Phys. Rev. Lett.} {\bf 71}, 3401 (1993). 
%
\bibitem{AKH} N. Hayashida {\em et al., Phys. Rev. Lett.} {\bf 73}, 
 3491 (1994). 
%
\bibitem{HillAnnRev} A.M.~Hillas, {\em Ann. Revs. Astr. Astrophys.},
 {\bf 22}, 425 (1984).
%
\bibitem{RB12} J.P.~Rachen and P.L.~Biermann, {\em Astron. Astrophys.}  
 {\bf 272}, 161 (1993); J.P.~Rachen, T.~Stanev and P.L.~Biermann, 
 {\em Astron. Astroph.} {\bf 273}, 377 (1993).
%
\bibitem{Arnold} S.S.~Al--Dargazelli {\em et al. J. Phys. G: Nucl. Part. Phys},
 submitted; J.~Szabelski, J.~Wdowczyk \& A.W.~Wolfendale,
 {\em J.~Phys.G: Nucl.Phys.} {\bf 12}, 1443 (1986). 
%
\bibitem{BhHSSL} P.~Bhattacharjee, C.T.~Hill and D.N. Schramm,
 {\em Phys. Rev. Lett.}, {\bf 69}, 567 (1992); G.~Sigl, D.N.~Schramm
 and P.~Bhattacharjee, {\it Astropart. Phys.} {\bf 2}, 401 (1994).
%
\bibitem{GRB1} E.~Waxman, {\em Phys. Rev. Letters}, {\bf 75}, 386 (1995).
%
\bibitem{GRB2} M.~Vietri, {\em Ap.~J.} {\bf 453}, 883 (1995).
%
\bibitem{GRB3} M.~Milgrom and V.~Usov, {\em Ap.~J.} {\bf 449}, L37 (1995).
%
\bibitem{FYSC} D.J. Bird {\em et al.} Phys. Rev. Lett. 71, 3401 (1993).
%
\bibitem{AGASAS} N.~Hayashida {\em et al. J. Phys. G: Nucl. Part. Phys.},
 {\bf 21} 1101 (1995).
%
\bibitem{WWSUGAR}X.~Chi, J.~Wdowczyk \& A.W.~Wolfendale, {\it J. Phys.
 G: Nucl. Part. Phys.} {\bf 18} (1992) and references therein.
%
\bibitem{PRL}T.~Stanev {\em et al., Phys. Rev. Lett.}, 
 {\bf 75}, 3056 (1995).
%
\bibitem{ASGP}N.~Hayashida {\em et al.} (The AGASA
 collaboration) {\em Phys. Rev. Lett}, submitted.
%
\bibitem{Adel} L.J.~Kewley, R.W.~Clay \& B.R.~Dawson, {\em Astropart. Phys.},
 in print.
%
\bibitem{HSV}  F. Halzen {\em et al., Astropart. Phys.} {\bf 3}, 151 (1995).
%
\bibitem{Auger}  See "Cosmic Rays Above 10$^{19}$~eV -- 1992", eds.
 M.~Boratav {\it et al.} {\it Nucl. Phys.} {\bf 28B} (1992); ``The Pierre
 Auger Project'', Design Report, The Auger Collaboration (1995).
%
\bibitem{GSH?} T.K.~Gaisser {\em et al., Phys. Rev.}, {\bf D43}, 314 (1991).
%
\bibitem{TP} T.J.L.~McComb, R.J.~Protheroe \& K.E.~Turver, {\em J. Phys. G:
 Nucl. Phys}, {\bf 5}, 1613 (979).
%
\bibitem{Ahhh} F.A.~Aharonian, B.L.~Kanewski \& V.V.~Vardanian,
 {\em Astrophys. Sp. Sci.}, {\bf 167}, 111 (1990).
%
\bibitem{Erber} T.~Erber, {\em Revs. Mod. Phys.}, {\bf 38}, 626 (1966).
%
%
\bibitem{McBLa}B.~McBreen \& C.J.~Lambert, {\em Proc. 17th Int. Cosmic Ray
 Conf.} (Paris) {\bf 6}, 70 (1981).
%
\bibitem{VS} H.P.~Vankov and P.V.~Stavrev, {\em Phys. Lett. B}, {\bf 266},
 178 (1991).
%
\bibitem{LPM} See T. Stanev {\em et al., Phys. Rev.} {\bf D25}, 1291 
   (1982) for the influence of the LPM effect on the shower development. 
%
\end{references}
\end{document}